\documentclass[twocolumn,amsmath,amssymb,floatfix,prl]{revtex4}
\usepackage{bm}
\usepackage{amsmath}
\usepackage{epsf}

\begin{document}

\title{Antimatter from the cosmological baryogenesis and the
anisotropies and polarization of the CMB radiation}
\author{Pavel D. Naselsky$^{1,2,3}$, Lung-Yih Chiang$^{1}$}

\affiliation{
$^{1}$Theoretical Astrophysics Center, Juliane Maries Vej 30, 2100
  Copenhagen, {\O} Denmark\\
$^{2}$Niels Bohr Institute, Juliane Maries Vej 30, 2100 Copenhagen,
{\O} Denmark \\
$^{3}$Rostov State University, Zorge 5, 344090 Rostov-Don, Russia}

\newcommand{\nc}{\newcommand}
\newcommand{\beq}{\begin{equation}}
\newcommand{\eeq}{\end{equation}}
\newcommand{\be}{\begin{eqnarray}}
\newcommand{\ee}{\end{eqnarray}}
\newcommand{\num}{\nu_\mu}
\newcommand{\nue}{\nu_e}
\newcommand{\nut}{\nu_\tau}
\newcommand{\nus}{\nu_s}
\newcommand{\mnus}{M_s}
\newcommand{\taus}{\tau_{\nu_s}}
\newcommand{\nnt}{n_{\nu_\tau}}
\newcommand{\rnt}{\rho_{\nu_\tau}}
\newcommand{\mnt}{m_{\nu_\tau}}
\newcommand{\tnt}{\tau_{\nu_\tau}}
\newcommand{\rar}{\rightarrow}
\newcommand{\lar}{\leftarrow}
\newcommand{\lrar}{\leftrightarrow}
\newcommand{\dm}{\delta m^2}
\newcommand{\mpl}{m_{Pl}}
\newcommand{\mbh}{M_{BH}}
\newcommand{\nbh}{n_{BH}}
\newcommand{\crit}{{\rm crit}}
\newcommand{\ini}{{\rm in}}
\newcommand{\cmb}{{\rm cmb}}
\newcommand{\rec}{{\rm rec}}

\newcommand{\Odm}{\Omega_{\rm dm}}
\newcommand{\Ob}{\Omega_{\rm b}}
\newcommand{\Om}{\Omega_{\rm m}}
\newcommand{\nb}{n_{\rm b}}
\def\simlt{\lesssim}
\def\simgt{\gtrsim}
\def\Cl{C_{\ell}}
\def\out{{\rm out}}
\def\in{{\rm in}}
\def\mean{{\rm mean}}
\def\zrec{z_{\rm rec}}
\def\zreio{z_{\rm reion}}
\def\wmap{{\it WMAP }}
\def\planck{{\it Planck }}
\begin{abstract}
We discuss the hypotheses that cosmological baryon asymmetry and
entropy were produced in the early Universe by phase transition of the
scalar fields in the framework of spontaneous baryogenesis scenario.
We show that annihilation of the matter-antimatter
clouds during the cosmological hydrogen recombination could distort of
the CMB anisotropies and polarization by delay of the
recombination. After recombination the annihilation of the
antibaryonic clouds (ABC) 
and baryonic matter can produce peak-like reionization at the high
redshifts before formation of quasars and early
galaxy formation. We discuss the constraints on the parameters of
spontaneous baryogenesis scenario by the recent \wmap CMB
anisotropy and polarization data and on possible manifestation of the
antimatter clouds in the upcoming \planck data.
\end{abstract}
\maketitle
\section{Introduction}
Recent release of the first-year \wmap data has confirmed that our
Universe is non-baryonic dominated. The vast
collection of stars, galaxies and clusters nevertheless contains a huge
amount of baryons without strong evidence of antibaryon
contamination to the spectrum of electromagnetic radiation in the
Universe. Does it mean that starting from the baryogenesis
epoch all antibaryons, or more generally
speaking, antimatter annihilate with the baryonic matter producing radiation
and only relatively small amount of antibaryons can survive
up to the present day during the expansion of the Universe? The answer
to this question has been the point of discussions in the literature
(see for review in
\cite{ds,kholov,jedamzik,dolgov}) including the Big Bang
Nucleosynthesis  properties,
antiprotons in the vicinity of the Earth and so on. The aim of this
paper is to investigate antimatter contamination in the recent CMB
data, namely, the \wmap anisotropy and polarization data through
distortions of the hydrogen recombination kinetics and possible late
reionization of the plasma and make the corresponding prediction for the
future \planck mission.

We re-examine the baryogenesis models following the arguments by \cite{ds}, in
which the baryonic and antibaryonic matter are very non-uniformly
distributed at very small scales  (for example, the corresponding mass
scale can be equivalent to $M \sim 10^3-10^5 M_{\odot}$ \cite{ds} and
follows the adiabatic perturbation upon these scales. Obviously, the
possibility of having non-uniformly distributed baryonic fraction of the
matter at very small scales is related to the Affleck-Dine
baryogenesis \cite{affleckdine} or the spontaneous baryogenesis mechanism
\cite{cohen}.
Taking into account the electromagnetic cascades driven by proton-antiproton
annihilation at the epoch of hydrogen recombination, we will show how
they distort the kinetics of the recombination producing corresponding features
in the CMB anisotropy and polarization power spectrum. Then we will discuss
possible late reionization of the hydrogen by the product of annihilation
and the corresponding transformation of the CMB anisotropy and
polarization power spectrum taking into account present \wmap and {\it
CBI} observational data. Finally we will show that the upcoming
\planck mission will be able to detect corresponding manifestation of
matter -antimatter annihilation even if the well-known
Sunyaev-Zeldovich $y$-parameter would be one order of magnitude
smaller that the {\it COBE} FIRAS limit \cite{fixen,mather,bersanelli}.

\section{Baryon-antibaryon bubble formation in the Universe}
It is assumed \cite{ds} that scalar baryon of SUSY model $\xi$ is coupled
to the scalar inflaton field $\Phi$ by the following potential
\begin{equation}
V_{int}(\xi,\Phi)=(\lambda\xi^2 +h.c.)(\Phi-\Phi_\crit)^2,
\label{eq1}
\end{equation}
where $\lambda$ is the coupling constant and $\Phi_\crit$ is some
critical value of the $\Phi$ field, which determines the point of
minimum of the $V_{int}(\xi,\Phi)$ potential. Starting from the high
values $\Phi_{int} \gg \Phi_\crit$ the
inflaton field decreases down to $\Phi_\crit$ and $V_\ini(\xi,\Phi)$
potential reach the point of minimum, while at $\Phi \ll \Phi_\crit$
for $V_{int}(\xi,\Phi)$ potential we will have $V_{int}(\xi,\Phi) =
(\lambda \xi^2 + h.c.)\Phi_\crit^2= V(\xi)$ independently on the
properties of the $\Phi$ field. It has been shown \cite{ds,kholov}
that because of the properties of the interactions the
most favorable conditions for baryogenesis might be created only for a
short time scale. It corresponds to a relatively small spatial
scales. Thus, the general picture of the baryonic matter-antimatter
spatial distribution would be similar to random distribution of the
islands with high baryon (or anti-baryon) asymmetry floating in the
the normal matter with $\beta=n_\cmb/n_b \simeq 5
\times 10^{-10}$, where $n_\cmb$ and $n_b$ are the present number
densities of the CMB photons and baryons. The mass distribution
function of the baryon (anti-baryon) clouds (ABC) is also estimated \cite{ds}
\begin{equation}
\frac{dn}{dM}\propto \exp\left[-\gamma^2
\ln^2\left(\frac{M}{M_\crit}\right)\right]
\label{eq2}
\end{equation}
where $\gamma$ and $M_\crit$ are free parameters of the theory.
As one can see from Eq.(\ref{eq2}), if $\gamma \gg 1$ then the mass
spectrum is localized at $M \sim M_\crit$, while for $\gamma \sim 1$
the mass spectrum will have monotonic character for the clouds
distribution over wide range of masses.
Dolgov and Silk \cite{ds} have also pointed out that $M_\crit$ could
be close to the solar mass $M_{\odot}$, but the range of  $M_\crit$
can be naturally expanded to  $10^3-10^5 M_{\odot}$
\cite{kholov}. Let us assume
that parameter $\gamma$ has especially high value: $\gamma \gg 1$ and
the initial distribution function of the baryon-antibaryon clouds is
close to the Dirac-$\delta$ function: $dn/dM \propto \delta_D (M
-M_\crit)$ and the characteristic size of clouds  $R_{cl} \propto
M^{1/3}_\crit$ is much
smaller than the size of the horizon $R_\rec$ at the
epoch of recombination ($z \simeq 10^3$): $R_{cl} \ll R_\rec$. We
denote $ \rho_{b,\in}$ and $\rho_{b,\out}$  the anti-baryon
density inside and baryon density outside the clouds, respectively, and the
mean
density $\rho_{b,\mean}$ at the scales much greater than $R_{cl}$ and
distances  between them,

\begin{equation}
\rho_{b,\mean}=\rho_{abc,\in} f + \rho_{b,\out}(1-f),
\label{eq3}
\end{equation}
where $f$ is the volume fraction of the clouds. We denote
\begin{equation}
\eta= \frac{\rho_{abc,\in}}{\rho_{b,\out}}.
\label{eq4}
\end{equation}

We can write down the following relations between the mean value of the
 density and inner and outer values
\begin{equation}
\rho_{b,\in}=\frac{\eta \rho_{b,\mean}}{1+f(\eta -1)},
\label{eq5}
\end{equation}
and
\begin{equation}
\rho_{b,\out}=\frac{ \rho_{b,\mean}}{1+f(\eta -1)}.
\label{eq6}
\end{equation}
Using the functions $f$ and $\eta$ we can define the anti-baryonic mass
fraction
\begin{equation}
F_b=\frac{\eta f}{1+f(\eta -1)},
\label{eq7}
\end{equation}
which is a function of the characteristic mass scale $M_0$ of the
anti-baryonic clouds.

Obviously, all the parameters $f$, $\eta$ and $F_b$ are the results
of the fine tuning of the inflaton $V_\in(\xi,\Phi)$ leading to
the formation of baryonic asymmetry in the Universe.

\section{Matter-antimatter baryonic clouds in the hot plasma}
At the end of inflation the Universe became radiation-dominated by mostly
light products of the inflaton decay. Some fraction of matter, however,
can exist with a form of primordial anti-baryonic clouds. Let us describe
the dynamics of such ABC evaporation
in the hot plasma.
  For
simplicity we will further assume that a single ABC has spherically
symmetric density distribution ($\rho_{in} \equiv \rho_{in}(r)$) with
the characteristic scale $R$ starting from which the contact between
ABC and the outer baryonic matter leads to energy release due to annihilation
\begin{equation}
\frac{dE}{dt}= 4\pi R^2\varepsilon_{\out}v_{T}
=4\pi R^2 c\varepsilon_{\out}\left(\frac{3 k T }{2 m_p
c^2}\right)^{\frac{1}{2}}
\label{eq8}
\end{equation}
where $v_{T}=\left(3kT/2m_p c^2 \right)^{1/2}$ is the speed
of sound in the plasma, $\varepsilon_{\out}$ is the energy density of the outer
plasma, $k$ is the Boltzmann constant, $m_p$ is the proton mass and $T$ is
the temperature of the outer plasma. Using Eq.(\ref{eq8}) and
the energy  of the inner ABC matter $E_{cl}=M_{cl}\,c^2 \sim (4\pi R^3/3)
\eta \varepsilon_{\out} $ for the characteristic time of evaporation we get
\begin{equation}
\tau_{ev}\simeq\frac{E_{cl}}{dE/dt}=\frac{\eta R}{3c}
\left(\frac{3kT}{2m_p c^2}\right)^{-1/2}.
\label{eq9}
\end{equation}
Equation (\ref{eq9}) indicates that any clouds with the size above
$R_{cr}\simeq (10^{-5} \div
10^{-4})\eta^{-1}\left(z/\zrec\right)^{1/2} r_h(\zrec)$ will survive up to the
moment of the cosmological hydrogen recombination
$t_{\rm rec}\simeq 2/3 (\Omega_m H_0^2)^{-1/2} \zrec^{-3/2}$, where
$\zrec \sim 10^3$ is the redshift of the recombination, $H_0=100h$ is the
present value of the Hubble constant, $\Omega_m$ is the baryonic plus
dark matter density scaled to the critical density and $r_h(\zrec)$ is
the horizon at the moment of recombination. The baryonic mass at the
moment of recombination is in order of magnitude $10^{19} M_{\odot}$
\cite{kt} and the corresponding mass scale of the ABC should be
roughly $(10^4 \div 10^7 M_{\odot}) \eta^{-3}$. If the $\eta$
parameter is close to unity, which means that density contrast between
the inner and outer zones is small, then the corresponding mass scale of the
ABC would be $10^4 \div 10^7 M_{\odot}$.
However, if $\eta \sim 10$, the corresponding mass scale of the ABCs
could be smaller, and comparable with the scale  $10 \div 10^4 M_{\odot}$.

\subsection{ABC at the nucleosynthesis epoch}
Let us compare the characteristic scales of the ABC with a few characteristic
scales of process in the framework of the Big Bang theory. Firstly,
the baryonic fraction of matter and its
spatial distribution play a crucial role starting from the epoch when the
balance between neutrinos ($\nu_e,\overline{\nu_e}$), neutrons ($n$)
and protons ($p$) in the following reactions $n + \nu_e \leftrightarrow p +
e^{-}$, $n + e^{+} \leftrightarrow p + \overline{\nu_e}$, $n \rightarrow p +
e^{-} + \overline{\nu_e}$ is broken. The corresponding time scale of
violation of the neutrino-baryon equilibrium is close to
$\tau_{\nu_e,p} \simeq 1$ sec when the temperature of the plasma was close to
$T_{\nu_e,p} \simeq 10^{10} $K (see
for the review in \cite{osw}). The time scale $\tau_{\nu_e,p}$ determines the
characteristic length $l_{\nu_e,p} \simeq c\tau_{\nu_e,p}$, which in
terms of the baryonic mass fraction of matter corresponds to
\begin{equation}
M_{\nu_e,p}\sim m_{pl}\left(\frac{\tau_{\nu_e,p}}{t_{pl}}\right)
\left(\frac{\rho_b}{\rho_{\gamma}}\right)|_{t=\tau_{\nu_e,p}}
\simeq 0.15 (\Omega_b h^2) M_{\odot},
\label{eq10}
\end{equation}
where $t_{pl}$ is the Planck time, $\rho_b$ and $\rho_{\gamma}$ are
the densities of baryons and radiation in the standard cosmological
model without anti-baryonic clouds. Following the SBBN theory we need to
specify the moment $\tau_{\rm end}$ when all light elements
(e.g. He$^4$ and deuterium) were synthesized during cosmological
cooling of the plasma. This moment is in order of the
magnitude close to $\tau_{\rm end} \sim 3\times 10^2 \div 10^3 $
sec. In term of the baryonic mass scale it corresponds to
\begin{equation}
M_{\rm end}\simeq
M_{\nu_e,p}\left(\frac{\tau_{\rm end}}{\tau_{\nu_e,p}}\right)^{3/2}
\simeq 5\times 10^3 \left(\frac{\tau_{\rm end}}{10^3 {\rm
sec}}\right)^{3/2} (\Omega_b h^2) M_{\odot}.
\label{eq11}
\end{equation}
Thus, if the characteristic mass scale $M_0$ for the baryonic clouds
is higher than $M_{\rm end}$, the cosmological nucleosynthesis within
each cloud and outside the clouds proceeds independently with others and
the mean mass fraction of each chemical element would be the same as in
SBBN theory. If all the anti-baryonic clouds will annihilate just before or
after hydrogen recombination  epoch , we will have simple renormalization of
the
baryonic matter density at the epoch of nucleosithesis
\begin{equation}
\rho_{b,\out}=\frac{ \overline{\rho_{b}}+\rho_{abc}f}{1-f}
\label{eq11}
\end{equation}
where $\overline{\rho_{b}}$ is the present day baryonic density rescaled to the
SBBN epoch. As one can see from Eq.(\ref{eq11}), if the fraction of the
ABC is small ($f\ll 1$), then all the deviation of the light-element mass
fractions from the SBBN predictions would be negligible.

\section{Energy release to the cosmic plasma from the ABC at the epoch
of hydrogen recombination}
The net of ABC produce the net of the high energy photons because of
annihilation at the boundary zones for each antimatter cloud. Using
Eq.(\ref{eq8}), we can estimate the rate of the energy injection to
the plasma as
\begin{equation}
\frac{d\varepsilon}{dt}=\frac{dE}{dt} n_{cl}=\frac{\rho_{cl}c^2}{\tau_{ev}}
\label{eq14}
\end{equation}
where $\rho_{cl}=M_{cl} n_{cl} $ and $n_{cl}$ is the spatial number
density of the ABC.
Let us define the mass fraction of the ABC as
$f_{abc}=\rho_{cl}/\rho_{\out}$ which determines the energy release to
the cosmic plasma at the epoch right before and during
hydrogen recombination. Because of Compton and bremsstrahlung interactions
the energy density of the products of annihilation leads to the CMB
energy spectrum distortion in different ways \cite{sz,ls}. If
$\tau_{ev}$ corresponds to the redshift $z > 3 \times 10^5
\left(\Omega_b h^2/0.022\right)^{-1/2}$ then we should get a
Bose-Einstein spectrum
\begin{equation}
n(x,\mu)=\left[\exp(x+\mu) -1\right]^{-1},
\label{eq14}
\end{equation}
where $x={\rm h}\nu/kT$ (here h is the Planck constant, not the Hubble
constant), $\nu$ is the frequency of the photons, $\mu$ is the
chemical potential:
\begin{equation}
\mu= \mu_0 \exp(-2x_0/x)
\label{eq15}
\end{equation}
where $x_0=0.018\left(\Omega h^2/0.125\right)^{7/8}$.
It has been shown \cite{sz} that chemical potential $\mu$ is related
with the energy release from annihilation by $\mu=3\rho_{abc} c^2/2
\varepsilon_r$, where $ \varepsilon_r= 4 \pi /c \int I(\nu) d\nu $
and $I(\nu)$ is intensity of the CMB.
For the redshift of annihilation below $z= 3\times 10^5\left(\Omega_b
h^2/0.022\right)^{-1/2}$ the distortions of the CMB power spectrum
follows to $y$-parameter type \cite{zeldovich}:
\begin{equation}
n(x)=\frac{1}{\sqrt{4\pi y}}\int d\xi\frac{\exp\left[-(\ln x +3y
-\xi^2)/4y\right]} {\exp(\xi)-1}
\label{eq16}
\end{equation}
where
\begin{equation}
y=\int_0^z\frac{k(T_e-T_{cmb})}{m_e c^2}\sigma_T n_e(z)c\frac{dt}{dz} dz
\label{eq17}
\end{equation}
and $\sigma_T$ is the Thomson cross-section, $n_e$ and $T_e$ are
 the electron number density and temperature, respectively.
The magnitude of $y$-distortion is related to the total energy transfer
by $\kappa = \Delta E/\varepsilon_r=\rho_{abc}
c^2/\varepsilon_r=\exp(4y)-1$. At the epoch $10^3 \le z \le 10^4$
the COBE FIRAS data give the constraint of the energy release from
annihilation $\kappa \le 2 \times 10^{-4}$, while $y \le 1.5 \times
10^{-5}$, and $\mu_0 \le 9 \times 10^{-5}$ at 95\% CL
\cite{fixen,mather,bersanelli}.

We would like to point out that the above mentioned properties of the
spectral distortions on the CMB power spectrum is based on the
assumption that the distribution of the anti-baryonic matter is
spatially uniform without any clusterization, and
therefore, no additional angular anisotropy  and polarization of the CMB
would have been produced during the epoch of hydrogen
recombination. However, cloudy structure of the spatial distribution
of anti-matter zones would generate spatial fluctuation
of the $y$-parameters, similar to the Sunyaev-Zeldovich effect
from the hot gas in clusters of galaxies at relatively higher redshift
$\sim \zrec$. Moreover, such clouds would produce
relatively higher but localized $y$-distortions on the CMB
power spectrum, which corresponds, in mean, to the COBE
FIRAS limit but locally could be much higher.

\section{Electromagnetic cascades and the hydrogen recombination}
As in previous section, below we want to estimate possible influence
of the electromagnetic products of annihilation on the ionization
balance at the epoch of
the hydrogen recombination. Using quantitative approach, we can assume that
because of the energy transfer for the photons from $E \sim m_p c^2$ down to
$E \sim I=13.6 $eV, where $I$ is the ionization potential, some
fraction $x_e \le 1$ could be reionized by the non-equilibrium quanta
from the  electromagnetic cascades in the plasma. The energy balance
for such ionization follows
\begin{equation}
I x_e n_{bar} \simeq \omega \varepsilon_r \kappa|_{z \sim \zrec}
\label{eq18}
\end{equation}
where $\omega$ is the efficiency of the energy transforms down to the
ionization potential range and $\zrec\simeq 10^3$.
>From Eq.(\ref{eq18}) one obtains
\begin{equation}
\omega \le 5.4 \times 10^{-6} \left(\frac{\Omega_b
h^2}{0.022}\right)\left(\frac{1+z}{1000} \right)^{-1}
\left(\frac{\kappa}{2 \times
10^{-4}}\right)^{-1}\left(\frac{x_e}{0.1}\right)
\label{eq19}
\end{equation}
Thus, the relatively small fraction ($\sim 10^{-5}$) of the
annihilation energy release can distort
the kinetics of the cosmological hydrogen recombination. The concrete
mechanism of the energy transition, starting from $E \simeq m_p c^2
\sim 1$ GeV down to $E \sim I$
is connected with the electromagnetic cascades of the annihilation
products with cosmic plasma. The annihilation of a nucleon and an
antinucleon produces  $\sim 5$ pions, 3 of which are charged
\cite{egidy}. For charged pions, electromagnetic cascade appears due
to ${\pi}^{(+,-)}\rightarrow  \mu^{(+,-)} +\nu^{(-,+)}_{\mu}$ decay
including $\mu^{(+,-)} \rightarrow e^{(+,-)} $ transition. The neutral
pions decay into two photons $\pi^{0} \rightarrow 2\gamma$. About 50\%
of the energy release is carried away by the neutrino, about 30\% by
the photons and about 17\% by electrons and positrons
\cite{steigman}. The spectrum of the decay has a exponential
shape $n(E)\propto \exp(-E/E_0)$,  where $E\ge E_0 \simeq 70$ MeV
\cite{egidy}. For the electron-positron pair and $\gamma$-quanta the
leading process of the energy redistribution down to ionization
potential are Compton scattering by the CMB photons and the
electron-positron pair production $\gamma + (H,He) \rightarrow (H,He)
+ e^{+} + e^{-}$. When $E\gg m_e c^2$, the Compton cross-section is
well approximated by the Klein-Nishina formula \cite{arons}
\begin{equation}
\sigma_C \simeq \frac{3}{8}\sigma_T\left(\frac{m_e
c^2}{E}\right)\left[\ln\left(\frac{2E}{m_e c^2}\right)+ \frac{1}{2} \right],
\label{eq20}
\end{equation}
where $\sigma_T$ is the Thomson cross-section. The corresponding
optical depth for the Compton scattering is $\tau_C \simeq 2.1
\sigma_T \left(m_e c^2/E_0\right) \simeq
7.5 \times 10^{-3} \tau_T$, where
\begin{equation}
\sigma_T= 56.7 \left(\frac{\Omega_b
h^2}{0.022}\right)\left(\frac{\Omega_m
h^2}{0.125}\right)^{-1/2} \left(\frac{1+z}{1000}\right)^{3/2}
\label{eq21}
\end{equation}
For the inverse Compton scattering of high energy electrons by the CMB
photons the corresponding optical depth is $\tau_{IC} \simeq 2 \times
10^9 \left(\frac{\Omega_b h^2}{0.022}\right)\tau_C\gg 1 $ for
$z \simeq 10^3$.

The pair production cross-section $\sigma_{pc}$ has the following
asymptotic for $w=E/m_e c^2>6$ \cite{zdziarski}
\begin{equation}
\sigma_{\rm He}\simeq 8.8 \alpha_f r^2_0 \ln\left(\frac{513 w}{825+w}\right)
\label{eq22}
\end{equation}
for neutral helium and
\begin{equation}
\sigma_{\rm H}\simeq 5.4 \alpha_f r^2_0 \ln\left(\frac{513 w}{825+w}\right)
\label{eq23}
\end{equation}
for the neutral hydrogen, where $r^2_0=\frac{3}{8\pi}\sigma_T$ and
$\alpha_f$ is the Fermi constant. Note that for ionized
hydrogen-helium which contain 76\% and 24\% of corresponding mass
fractions of the light elements, the optical depth is close to
\begin{equation}
\tau_{pc} \simeq 2.3 \left(\frac{\Omega_b
h^2}{0.022}\right)\left(\frac{\Omega_m h^2}{0.125}\right)^{-1/2}
\left(\frac{1+z}{1000}\right)^{3/2},
\label{eq24}
\end{equation}
for $w z \gg 825$ \cite{zdziarski}.

Thus, as one can see from Eq.(\ref{eq21})-(\ref{eq24}), the energy
loss for high-energy electrons is determined by the inverse Compton
scattering off the CMB photons, whereas for the high-energy photons
the main process of the energy loss is the electron-positron pair
creation by neutral and ionized atoms.

For the non-relativistic electrons ($w<1$) the optical depth inverse
Compton scattering is given by
$\tau_{IC}\simeq 2 \times 10^9 \tau_T$, whereas for the photons it
is close to the Thomson optical depth.
It has been shown \cite{arons,zdziarski} that for high energy
$\rightarrow$ low energy photons conversion the spectral number density is
\begin{equation}
\frac{dn(E)}{dE} \simeq \frac{A}{\sigma_T n_e c}\left(w^{-2}
+\frac{14}{5} w^{-1}\right)
\label{eq25}
\end{equation}
for $E < E_0$, which corresponds to energy density
\begin{equation}
\epsilon=\int E dE \frac{dn(E)}{dE} \simeq \frac{14A m_e c E_0}{5 n_e
\sigma_T} \left[1+\frac{5}{14} \left(\frac{m_e c^2}{E_0}\right)\right].
\label{eq26}
\end{equation}
Therefore, from Eq.(\ref{eq25})-(\ref{eq26}) we can estimate the
spectral energy density at the range $E \simeq I$
\begin{equation}
\epsilon(E \simeq I) \simeq \frac{5}{14} \ln2 \cdot
\epsilon\frac{m_e c^2}{E_0} \simeq 1.7 \times 10^{-3} \epsilon,
\label{eq27}
\end{equation}
which is much higher than the limit from Eq.(\ref{eq19}). Note that an
additional factor $0.47$ results from the fraction
of the annihilation energy related to electromagnetic component.
$\omega \simeq 8 \times 10^{-4}$.
As one can see, the non-equilibrium ionization of the primordial
hydrogen and helium at the epoch of recombination is more effective
than the distortions of the CMB blackbody power spectra.

\section{Distortion of the recombination kinetics}
The model of the hydrogen-helium recombination process affected by the
annihilation energy release can be described phenomenologically in
terms of the injection of additional $Ly_{\alpha}$ and $ Ly_c$ photons
\cite{psh,dn,doroshkevich}. For the epochs of antimatter clouds
evaporation ($\eta-1\ll 1$) the rate of ionized photon production
$n_{\alpha}$ and $n_{c}$ are defined as
\begin{eqnarray}
\frac{dn_{\alpha}}{dt} = \varepsilon_{\alpha}(t) \langle \nb(t)\rangle
H(t) , \nonumber\\
\frac{dn_{i}}{dt} = \varepsilon_{i}(t) \langle \nb (t)\rangle
H(t),
\label{eq28}
\end{eqnarray}
where $H(t)$ and $\langle\nb(t)\rangle$ are the Hubble parameter
and the mean baryonic density, respectively,
$\varepsilon_{\alpha,i}(t)$ are the efficiency of the
$Ly_{\alpha}$ and $ Ly_c$ photon production.
As one can see from Eq.~(\ref{eq28}) the dependence of
$\varepsilon_{\alpha,i}(t)$ parameters upon $t$ (or redshifts $z$)
allows us to model any kind of ionization regimes. For the ABC from
Eq.(\ref{eq19})-(\ref{eq20}) we have
\begin{equation}
\varepsilon_{\alpha,i} \simeq
\omega \left(\frac{m_pc^2}{I}\right)\left[H(t) \tau_{ev}\right]^{-1} f_{abc}.
\label{eq29}
\end{equation}
If the time of evaporation is comparable with the Hubble time
$H^{-1}(t)$ at the epoch of
recombination $z\sim \zrec$, then $ \varepsilon_{\alpha,i}$
parameters are constant and proportional to $f_{abc}$.

\begin{figure}[!h]
\epsfxsize=8 cm
\epsfbox{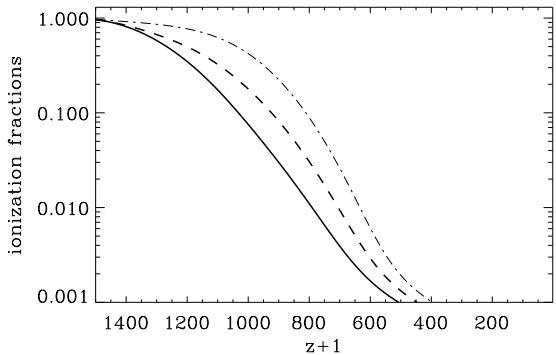}
\vspace{0.5cm}
\caption{ The ionization fractions for the model 1 (the solid line),
the model 2 (dash line) and model 3 (dash-dot line) as a function of
redshift.}
\label{if}
\end{figure}

We demonstrate the effectiveness of our phenomenological approach in
Fig.~\ref{if}: the ionization fraction $x_e$ against redshift for the three
models listed below:
\begin{itemize}
\item model 1: $\varepsilon_{\alpha}\simeq\varepsilon_i =1$;
\item model 2:$\varepsilon_{\alpha}\simeq\varepsilon_i=10$;
\item model 3: $\varepsilon_{\alpha}\simeq\varepsilon_i =100$.
\end{itemize}
The curves are produced from the modification of the {\sc recfast} code
\cite{recfast}. For all models we use the following values of the
cosmological parameters: $\Ob h^2=0.022,\Om h^2=0.125$,
$\Omega_{\lambda}=0.7$, $h=0.7$, $\Om + \Omega_{\lambda}=1$,
$H(t)\tau_{ev}\sim 1$.

As one can see from Fig.~\ref{if} all the models 1-3 produce delays of
recombination and can distort of the CMB anisotropy and polarization
power spectrum, which we will
discuss in the following Section. We would like to point out that our
assumption about the characteristic
time of the ABC evaporation, namely $H(t_{\rm rec})\tau_{ev} \sim 1$
implies that at $t\gg t_{\rm rec}$
all the ABC disappear. If $H(t_{\rm rec})\tau_{ev}
\gg  1$, however, at the epoch of recombination the corresponding
influence of the non-equilibrium photons can be
characterized by the renormalization of the $\varepsilon_{\alpha,i}$
parameters in the following way: $\varepsilon_{\alpha,i}(z) =
\varepsilon_{\alpha,i}(\zrec) (H(z)\tau_{ev})^{-1}$
where $\varepsilon_{\alpha,i}(\zrec)$ corresponds to the models
1-3. The mean factor, which should
necessarily be taken into account, is the absorption of the high energy
quanta from annihilation by the CMB photons. If, for example,
$\tau_{ev}$ corresponds to the redshift $z_{\rm rei} \sim 100$, then
\begin{equation}
\varepsilon_{\alpha,i}(\zrec)\simeq
\varepsilon_{\alpha,i}(\zrec)\left(\frac{z_{\rm reion}}{z_{\rm
rec}} \right)^{3/2} \sim 0.03\varepsilon_{\alpha,i}(z_{\rm rec}).
\label{eq30}
\end{equation}

For the relatively early reionization of the hydrogen by the products
of annihilation, the ionization
fraction of matter $x_e=n_e/\langle \nb\rangle$ can be obtained
from the balance between the recombination and the ionization process
\begin{equation}
\frac{dx_e}{dt}=-\alpha_{\rm rec}(T)\langle\nb\rangle x^2_e +
\varepsilon_{i}(z) (1-x_e)H(z)\Theta(z_{ev}-z),
\label{eq31}
\end{equation}
where $\alpha_{\rm rec}(T)\simeq 4\times 10^{-13} \left(T/10^4
K \right)^{-0.6}$ is the recombination coefficient ,$z_{ev}$
corresponds to $\tau_{ev}$ and $T$ is the
temperature of the plasma and $\langle\nb\rangle=n_b$ is the mean
value of the baryonic number density of the matter. In an equilibrium
between the recombination and the ionization process the ionization
fraction of the matter follows the well-known regime
\begin{equation}
\frac{x^{2}_e(z)}{1-x_e(z)} = \frac{\varepsilon_{i}(z) H(z)}{
\alpha_{\rm rec}(z) \nb(z)}\Theta(z_{ev}-z),
\label{eq32}
\end{equation}
where $H(z)=H_0\sqrt{\Om(1+z)^3+1-\Om}$
and $\nb \simeq 2 \times 10^{-7}(\Ob h^2/0.02)(1+z)^3$.
We would like to point out that Eq.(\ref{eq32}) can be used for any
models of the late reionization,
choosing the corresponding dependence of the $\varepsilon_{i}(z)$
parameter on redshift. This point is vital in our modification of the
{\sc recfast} and the {\sc cmbfast} packages, from which we can use the
standard relation for matter temperature $T(z)\simeq 270
\left(1+z/100 \right)^2 {\rm K}$ and
all the temperature peculiarities of the reionization and clumping would be
related with the $\varepsilon_{i}(z)$ parameter through the mimicking
of ionization history \cite{naselsky,late}.

From Eq.(\ref{eq32}) one can find the maximal value of the ionization
fraction at the moment $z\simeq z_{ev}$
\begin{equation}
x^{\rm max}_e=-\frac{1}{2} {\Gamma}+\left(
1+\frac{1}{4}{\Gamma}^2\right)^{1/2}
\label{eq33}
\end{equation}
where ${\Gamma}=\varepsilon_{i}(z_{ev}) H(z_{ev})/[
\alpha_{\rm rec}(z_{ev}) \nb(z_{ev})]$. At $10 \ll z< z_{ev}$ the
relaxation of the matter temperature to the CMB temperature proceeds
faster than the ionized hydrogen becoming neutral and for $x_e$ from
Eq.(\ref{eq31}) we get
\begin{equation}
x_e(t)\simeq x^{\rm max}_e\left(1+ x^{\rm
  max}_e\int\limits_{\tau_{ev}}^t  \alpha(T) \nb
dt\right)^{-1}.
\label{eq34}
\end{equation}
While the temperature of
matter is close to the CMB temperature $T_{\rm CMB}$, the
corresponding time of recombination is
\begin{equation}
\Delta t_{\rm rec}\simeq \frac{x_e}{|dx_e/dt|} \simeq
({x^{\rm max}_e})^{-1} t_{\rm rec}(T_{\rm CMB}),
\label{eq35}
\end{equation}
where   $t_{\rm rec}=[\alpha(T)\, \nb]^{-1} \ll \tau_{ev}, H^{-1}(t)$.

\begin{figure}[!h]
\epsfxsize=8 cm
\epsfbox{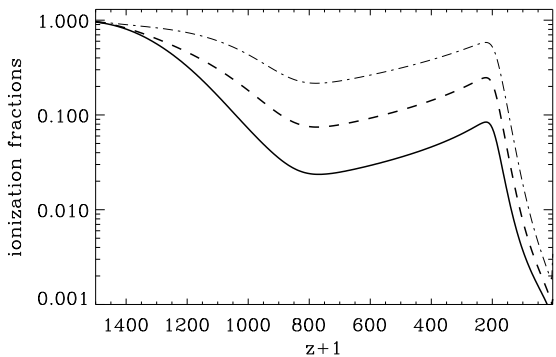}
\vspace{0.5cm}
\caption{
The ionization fractions for the model 4 (the solid line), the model
5(dash line) and model 6  (dash-dot line) as a function of redshift.}
\label{iff}
\end{figure}

In addition to the models 1-3 we introduce the following three models:
\begin{itemize}
\item model 4: $\varepsilon_{\alpha}\simeq\varepsilon_i=0.1\times
  \left[(1+z)/1000\right]^{3/2}$;
\item model 5: $\varepsilon_{\alpha}\simeq\varepsilon_i=1\times
  \left[(1+z)/1000\right]^{3/2}$;
\item model 6: $\varepsilon_{\alpha}\simeq\varepsilon_i=10\times
  \left[(1+z)/1000\right]^{3/2}$.
\end{itemize}
where $z_{ev}=200$. In Fig.\ref{iff} we plot the ionization fraction
for the models 4-6 versus redshift. As one can see from the
Fig.\ref{iff} the delay of recombination at $z=10^3$ is smaller than
in Fig.\ref{if}, but the reionization appears at $z \simeq z_{ev}$. At
the range of redshifts $z \gg z_{ev}$ the behavior of ionization
fraction follows Eq.(\ref{eq34}) with rapid decrease. The properties
of the models 4-6 are similar to those of the peak-like reionization
model \cite{late}.

\section{The CMB anisotropy and polarization features from the
  matter-antimatter annihilation}
In order to find out how sensitive is the polarization
power spectrum to the annihilation energy release, we consider
phenomenologically different variants of hydrogen reionization
models by modifying the {\sc cmbfast} code for the models 1-6
\cite{cmbfast}. One addition problem appears if we are interested in
observational constraints on the anti-matter fraction abundance related to
the late reionization of hydrogen at low redshift $z<20$. After
\wmap mission the most preferable value of the optical depth of reionization
is $\tau_{\rm reion}\simeq 0.17$ \cite{wmapresults}, while it is also shown
\cite{late} that even the ``standard model'' with $z_{\rm reion}=6$ is not
ruled out from the \wmap data (see also \cite{cfw}). Recently it has
been argued that the late reionization
could exist with two stages, one at $z_{\rm reion} \simeq 15$ and
$\zreio \simeq 6$ due to energy release from different population
of stars \cite{cen} or heavy neutrinos \cite{hansen}. Without
measurements with higher accuracy of the CMB polarization and
temperature-polarization cross-correlation, it is unlikely to settle
the issue on late reionization, even for \wmap resolution and sensitivity.
However, any assumption about the optical depth of the late
reionization are crucial for the estimation of any constraints on the
ABC abundance. If, for example, we adopt the \wmap limit $\tau_{\rm
reion} \simeq 0.17$ from the pure late reionization, the
peak-like or delayed recombination models from the ABC would by
restricted very effective.
But, if we assumes that roughly $\tau_{\rm reion} \sim 0.04$ comes
from late reionization and $\tau_{\rm reion} \sim 0.06 \div 0.12$ is
related to the ABC contamination at relatively high redshifts, then
the constraints on the ABC abundance would be rather smaller than for
the previous case. For estimation of the ABC features in the
CMB anisotropy and polarization power spectrum we use a
more conservative limit on the optical depth of reionization
$\tau_{\rm reion} \sim 0.04$ at  $\zreio \simeq 6$ in order to
obtain the upper limit on the ABC manifestation in the CMB data.
\begin{figure}[!h]
\epsfxsize=8 cm
\epsfbox{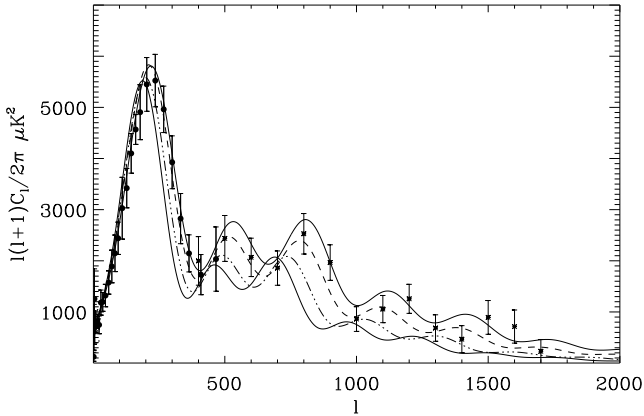}
\vspace{0.5cm}
\caption{
The CMB power spectrum for the standard model without energy
injection,  (the solid line),  the model 1(the dash line), model 2
(the dash-dot line) and model 3 (the lowest thick solid line) as a
function of redshift. For $\ell < 500$ we use the \wmap data \cite{wmappowerspectrum}, while
for $\ell > 500$ together with error bars the data is from {\it CBI}
experiment \cite{cbi}.}
\label{iff}
\end{figure}

\begin{figure}
\epsfxsize=8 cm
\epsfbox{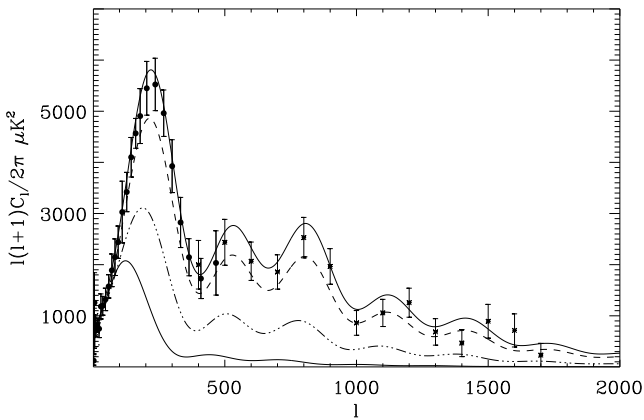}
\vspace{0.5cm}
\caption{
The CMB power spectrum for the standard model without energy injection
(the solid line), the model 4(dash line), 5 (dash-dot line) and 6 (the
lowest thick solid line) as a function of redshift.
The experimental data points are the same as in Fig.\ref{iff}.}
\label{iffa}
\end{figure}

\begin{figure}
\epsfxsize=8 cm
\epsfbox{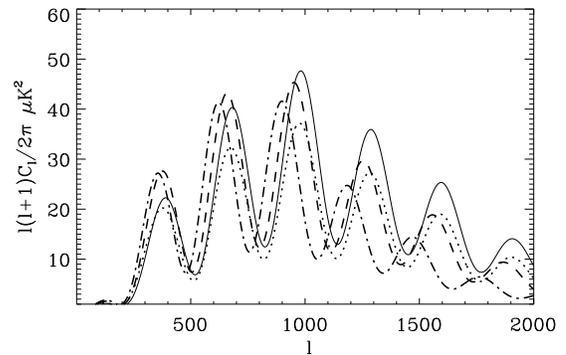}
\vspace{0.5cm}
\caption{The polarization power spectrum for the standard model (the
 solid line), the model 4 (the dot line),
the model 2 (the dash line) and the model 3 (the dash-dot line) as a
 function of redshift.}
\label{iffa1}
\end{figure}

In Fig.~\ref{iffa1} we plot the polarization power
spectrum $C_p(\ell)$ for the model 1 -6 plus the standard single
reionization model at $\zreio \simeq 6$ . The difference
between model 1 and 2 mainly lies in the multipoles $2< \ell <
30$.

\begin{figure}[!h]
\epsfxsize=8 cm
\epsfbox{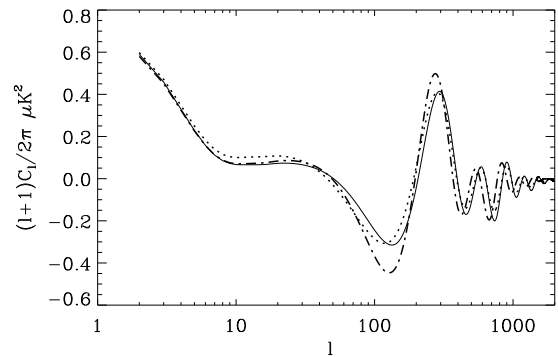}
\vspace{0.5cm}
\caption{The TE cross-correlation power spectrum for the models listed in
Fig.\ref{iffa1} with the same notations.}
\label{iffb}
\end{figure}

As one can see from the Fig.\ref{iffa},  in order of the magnitude the
$\varepsilon_{\alpha,i}$ parameters should be smaller than unity, if
$z_{ev} \simeq z_{\rm rec}$ and $\varepsilon_{\alpha,i}< 10^{-2}$,
if $z_{ev} \simeq 200$. So, using Eq.(\ref{eq29}) one can find that
\begin{equation}
f_{abc}=\omega^{-1}\varepsilon \left(H(t)\tau_{ev}\right)\frac{I}{m_p
 c^2}\le 1.7 \times 10^{-5} \left(\frac{1+z_{ev}}{200}\right)^{3/2},
\label{eq37}
\end{equation}
while from the spectral distortion of the CMB blackbody power spectra
we obtain
\begin{equation}
f^{y}_{abc}\le 1.7 \times 10^{-4}\left(\frac{\Omega_b
  h^2}{0.022}\right)\left(\frac{1+z_{ev}}{200}\right)
\label{eq38}
\end{equation}

\section{How \planck data can constrain the mass fraction of the antimatter?}
As is mentioned above, the observational constraint on the
antimatter mass fraction $f_{abc}$ depends on the accuracy of the
power spectrum estimation from the contemporary and upcoming CMB data sets.
As an example, how the upcoming \planck data would be important for
cosmology, we would like to compare the upper limit on the $f_{abc}$
parameter, using the \wmap and {\it CBI} data with the expected
sensitivity of the \planck data. We assume that all the systematic
effects and foreground contaminations should
be successfully removed and the accuracy of the $C_\ell$ estimation
would be close to the cosmic variance limit at low multipoles for both
the temperature anisotropies, polarization and the TE
cross-correlation as well.

The differences between the delayed recombination and early reionized
universe models in comparison with the expected sensitivity of the
\planck experiment can be
expressed in terms of the power spectrum $C^{a,p}(\ell)$
(for the anisotropy, and the $E$ component of polarization) \cite{doroshkevich}
\begin{equation}
 D^{a,p}_{i,j}(\ell) =
 \frac{2\left[C^{a,p}_{i}(\ell)-C^{a,p}_{j}(\ell)\right]}{C^{a,p}_{i}
 (\ell)+ C^{a,p}_{j}(\ell)},
\label{eqd}
\end{equation}
where the indices $i$ and $j$ denote the different models and $a$ and $p$
denote anisotropy and polarization.
\begin{figure}[!h]
\epsfxsize=8 cm
\epsfbox{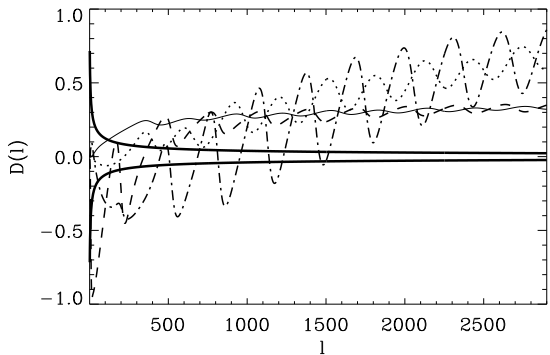}
\vspace{0.5cm}
\caption{ The $ D^{a,p}_{i,j}(\ell)$ function for different models of
ionization. The solid line corresponds to $ D^{a}_{i,j}(\ell)$ for $i=0$(
standard model without the ABC) and $j=4$; the dot line is
$ D^{a}_{i,j}(\ell)$ for $i=0$ and $j=1$; the dash line corresponds to
$ D^{p}_{i,j}(\ell)$ for $i=0$ and $j=4$; the dash-dot line is
$ D^{p}_{i,j}(\ell)$ for $i=0$ and $j=1$. The thick solid lines represent
the error bars limit from cosmic variance.}
\label{diff}
\end{figure}
In order to clarify the manifestations of the complex
ionization regimes in the models 1 and 4 we need to compare the peak
to peak amplitudes of the $D^{a,p}_{i,j}(\ell)$ function with the
expected error of the anisotropy power spectrum for \planck
experiment. We assume that
the systematics and foreground effects are successfully removed. The
corresponding error bar should be
\begin{equation}
\frac{\Delta C_{\ell}}{C_{\ell}} \simeq \frac{1}{\sqrt{f_{\rm
sky}(\ell+\frac{1}{2})}}\left[1+ w^{-1}C_{\ell}^{-1} W^{-2}_{\ell}\right],
\label{c}
\end{equation}
where $w=(\sigma_{p}\theta_{\rm FWHM})^{-2}$, $W_{\ell}\simeq
\exp\left[-\ell(\ell+1)/2\ell^2_s\right]$,
$f_{\rm sky} \simeq 0.65$ is the sky coverage during the first year of
observations, $\sigma_{p}$ is the sensitivity per resolution element
$\theta_{\rm FWHM} \times \theta_{\rm FWHM}$ and $\ell_s=\sqrt{8\ln 2}
\, \theta^{-1}_{\rm FWHM}$.

As one can see from Fig.~\ref{diff} for $ D^{a,p}_{i,j}(\ell)$ the
corresponding peak to peak amplitudes are at the order of magnitude
$5 \div 10$ times higher than the error bars limit at $\ell \sim
1500\div2500$. That means that both anisotropies and the polarization power
spectra caused by the complicated ionization regimes can be tested
directly for each multipole of the $\Cl$ power spectrum
by \planck mission, if the systematic effects are removed down to the
cosmic variance level. Moreover, at the 95\% CL the corresponding
constraint on the $f_{abc}$ parameter can be
$2.5\div 5$ times  smaller than the limit from Eq.(\ref{eq37}), or in
principle, the upcoming \planck mission should be able to detect any
peculiarities caused by the antimatter annihilation during the epoch
of the hydrogen recombination.

It is worth noting that in this paper we do not discuss the direct
contribution of antimatter regions to the CMB anisotropy formation,
assuming that their corresponding size is smaller than the typical
galactic  scales, and also smaller than the corresponding angular
resolution of the recent CMB experiments such as {\it WMAP}, {\it
  CBI}, {\it ACBAR}.  If the size of the ABC is
comparable with the size of galactic or cluster scales, they could manifest
themselves as point-like sources in the CMB map. For the upcoming
Planck mission there are well defined predictions for the number
density of bright point sources for each frequency band at
the range $30 \div 900$ GHz. It would be interesting to obtain a new
constraint on the ABC fraction for large-scale clouds. This work is
in progress.

\smallskip
{\it Acknowledgments:} This paper was supported in part by Danmarks
Grundforskningsfond through its support for the establishment of TAC.

\vfill

%
%
\newcommand{\autetal}[2]{{#2.\ #1 {\it et al}.,}}
\newcommand{\aut}[2]{{#2.\ #1,}}
\newcommand{\saut}[2]{{#2.\ #1,}}
\newcommand{\laut}[2]{{and #2.\ #1,}}
\newcommand{\psaut}[2]{{#2.\ #1}}

%
%
\newcommand{\refs}[6]{#2 {\bf #3}, #4 (#5).}
\newcommand{\midrefs}[6]{#2 {\bf #3}, #4 (#5);}
\newcommand{\unrefs}[6]{#2 #3 (#6).}
\newcommand{\midunrefs}[6]{#2, #3 (#6);}

%
\newcommand{\book}[6]{{\it #1} (#2\, #5).} 
\newcommand{\midbook}[6]{{\it #1} (#2\, #5);} 

%
\newcommand{\proceeding}[6]{#5, in #4, edited by #3 (#2#5).} 
\newcommand{\midproceeding}[6]{#5, in #4, edited by #3 (#2#5);} 

\newcommand{\combib}[3]{\bibitem{#3}} 

%
%
\def\apjl{Astrophys.\ J.\ Lett.}
\def\mn{Mon.\ Not.\ R.\ Astron.\ Soc.}
\def\nature{Nature (London)}
\def\araa{Ann. Rev. Astron. Astrophys.}
\def\aa{Astron.\ Astrophys.}
\def\apj{Astrophys.\ J.}
\def\apjs{Astrophys.\ J. Supp.}
\def\prl{Phys.\ Rev.\ Lett.}
\def\prd{Phys.\ Rev.\ D}
\def\pl{Phys.\ Lett.}
\def\np{Nucl.\ Phys.}
\def\rmp{Rev.\ Mod.\ Phys.}
\def\cmp{Comm.\ Math.\ Phys.}
\def\mpl{Mod.\ Phys.\ Lett.}

\newcommand{\amp}{\& }
\def\cqg{Class.\ Quant.\ Grav.}
\def\grg{Gen.\ Rel.\ Grav.}

\def\cambridgepress{Cambridge University Press, Cambridge, UK}
\def\princetonpress{Princeton University Press}
\def\worldpress{World Scientific, Singapore}
\def\oxfordpress{Oxford University Press}

\end{document}